# Wave localization at the boundary of disordered photonic lattices

A. Szameit[1], Y. V. Kartashov[2], P. Zeil[3], F. Dreisow[3], M. Heinrich[3], R. Keil[3], S. Nolte[3], A. Tünnermann[3], V. A. Vysloukh[2], and L. Torner[2]

[1]*Physics Department and Solid State Institute, Technion, 32000 Haifa, Israel*

[2]*ICFO-Institut de Ciencies Fotoniques, and Universitat Politecnica de Catalunya, Mediterranean Technology Park, 08860 Castelldefels (Barcelona), Spain*

[3]*Institute of Applied Physics, Friedrich Schiller University Jena, Max-Wien-Platz 1, 07743 Jena, Germany*

We report on the experimental observation of reduced light energy transport and disorder-induced localization close to a boundary of a truncated one-dimensional (1D) disordered photonic lattice. Our observations uncover that near the boundary a higher level of disorder is required to obtain similar localization than in the bulk.

Light and particle dynamics in disordered media is a topic of continuously renewed interest. In solid-state physics, particles scattered by lattice defects generate a random walk, resulting in a transition from ballistic transport to complete suppression of transport when disorder exceeds certain level [1]. Due to this transition, the infinitely extended eigenmodes of the system are transformed into localized modes [2]. This is a universal concept applicable to a variety of physical settings [3,4]. Optical setups proved to be promising candidates to observe localization in random media [5-10] due to analogy of a solid and photonic lattice, where longitudinally invariant disorder can be realized [11,12]. Since the localization relies on fluctuations imposed on otherwise periodic structure, a truncation of a disordered lattice yields distortions of the underlying periodicity, which may have a strong impact on wave localization.

In this Letter we observe wave-packet localization at the edge of disordered truncated 1D lattices for sufficiently large levels of disorder. A higher degree of disorder (as compared to disorder required for localization in the lattice center) is required to achieve the same degree of localization at the surface of the lattice, which we attribute to a repulsive potential arising in the vicinity of the boundary [13].

In the tight-binding approximation, the field $q_\mu$ in the $\mu$-th guide ($\mu \geq 0$) after exciting the boundary waveguide ($\mu = 0$) is given by $q_\mu = i^\mu[J_\mu(2c\xi) + J_{\mu+2}(2c\xi)]$ [14], where $c$ is the coupling constant, $\xi$ is the propagation distance, and $J_\mu(2c\xi)$ is the $\mu$-th order Bessel function. The first term here corresponds to the solution in the infinite lattice, whereas the second one appears due to reflection at the boundary. Rewriting this expression as $q_\mu = i^\mu[(\mu+1)(c\xi)^{-1}]J_{\mu+1}(2c\xi)$ reveals that the side lobes of the discrete diffraction pattern [15], described by the first maximum of the Bessel functions [16], are slightly shifted to larger $\xi$ values (as compared to the infinite array) since the order of Bessel function is increased by one. Hence, the effective diffraction upon excitation of the boundary guide is smaller than diffraction in infinite array, although the relative difference in side-lobe positions is vanishing for $c\xi \gg 1$. In contrast, the amplitude $q_0 = (c\xi)^{-1}J_1(2c\xi)$ in the excited edge guide for $c\xi \gg 1$ decreases much faster with distance as compared to the excited guide in an infinite array, due to the factor $(c\xi)^{-1}$. Hence, the boundary is repulsive [16,13] and, in consequence, a higher disorder level is required to overcome delocalization in the edge waveguide.

In our simulations, we employ a continuous Schrödinger equation to describe the evolution of the field $q$ in a disordered array:

$$i\frac{\partial q}{\partial \xi} = -\frac{1}{2}\frac{\partial^2 q}{\partial \eta^2} - pR(\eta)q. \qquad (1)$$

Here $\eta, \xi$ are the normalized transverse and longitudinal coordinates, $p$ is the refractive index modulation depth, $R(\eta) = \sum_{m=-(M-1)/2}^{(M-1)/2} \exp[-(\eta - \eta_m - md)^6/W^6]$ is a random function describing the refractive index profile, $d$ is the average spacing, $\eta_m < d/2$ is a random shift of $m$-th guide center (uniformly distributed in $[-S_d, +S_d]$), $W$ is the waveguide width, and $-15 \leq m \leq 15$ (31 waveguides in each array). We adapted a Monte-Carlo approach when in each realization $\eta_m$ was calculated and refractive index profile $R(\eta)$ was constructed. The propagation of single-site Gaussian beam up to the distance $L$ was simulated with a split-step Fourier method. We set $W = 0.3$ (waveguide width of 3 $\mu$m), $d = 1.6$ (waveguide spacing 16 $\mu$m), $L = 69$ (100 mm of propagation), $p = 11$ (refractive index modulation $\sim 7.6 \times 10^{-4}$). The normalized disorder parameter $S_d^{\text{norm}}$ was varied from 0 (regular array) up to 0.7 (strongly disordered array with $S_d = 7$ $\mu$m). A statistical averaging of the intensity distributions for 1000 lattice realizations provides the information about localization of the light energy. The key issue was the comparison of localization for the excitation of the central ($m = 0$) and the edge ($m = \pm 15$) guides for different disorder levels $S_\text{d}$. Figure 1 illustrates the averaged output intensity pattern (in log scale) for excitation of the central and edge guides. At low disorder [$S_\text{d} = 3$ $\mu$m, Fig. 1(a)] the averaged output intensity represents a superposition of a triangular distribution and two maxima which result from the side lobes of the discrete diffraction in regular systems. With increasing disorder the positions of those two maxima remain almost unchanged, but their amplitudes decrease, so that at high disorder [$S_\text{d} = 7$ $\mu$m, Fig. 1(b)] a perfect triangular distribution forms indicating on exponential localization.

This behavior is confirmed experimentally in fs-laser-written waveguide arrays in fused silica [17]. The light at $\lambda = 633$ nm was launched into a single guide using a 20× microscope objective and projected on a camera using a 4× objective. To collect statistics we fabricated 30 different waveguide arrays for each disorder level $S_\text{d} = 0,1,...,7$ $\mu$m. Averaged output intensity patterns are shown in Fig. 2. At $S_\text{d} = 0$ $\mu$m the profile for the homogeneous array is obtained [Fig. 2(a)]. Increasing the disorder to $S_\text{d} = 2$ $\mu$m yields a partial localization [Fig. 2(b)] for both center and edge excitation. Strong disorder ($S_\text{d} = 7$ $\mu$m) results in localization (exponential) of light in the proximity of the excited guide [Fig. 2(c)].

While qualitative features of light localization for center and boundary excitations are similar, the *degree* of localization is different for a fixed $S_\text{d}$. To show this, we compare the slopes $\alpha_\text{c}, \alpha_\text{s}$ (subscripts c and s pertain to excitations of central and surface guides, respectively) of the inner linear part of the output averaged $\ln(I)$ distributions (characterizing the exponential decay rates of the averaged $I(\eta)$ dependencies) as a function of $S_\text{d}$. Simulations are depicted in Fig. 3(a), which confirms the experimental data [Fig. 3(b)]. The graphs start at a disorder level of $S_\text{d} = 2$ $\mu$m to ensure the triangular shape of $\ln(I)$ distribution. According to the exponential decay rate criterion, for a fixed disorder level the localization in the center is always stronger than at the boundary. We attribute this to the repulsive action of the boundary. This additional delocalizing factor has to be overcome by disorder to achieve similar localization as in the case of excitation of a central guide. The standard deviation of the decay rates reduces monotonically with increasing $S_\text{d}$ [$S_\text{d} = 3$ $\mu$m : 16% (theory), 20% (experiment); $S_\text{d} = 7$ $\mu$m : 2.5% (theory), 4% (experiment)].

The features of localization are closely related with the transformation of eigenfunctions of disordered lattice. In an infinite system all eigenmodes are localized even for minimal disorder. However, since we deal with a finite sample, we distinguish between strongly and weakly localized (extended) modes. In Fig. 4, examples of calculated fundamental eigenmodes of a disordered array that are localized close to the lattice center (left column) and boundary (right column) are shown for different disorder levels. A progressive localization with increase of disorder is apparent. The evolution of the input beam at low disorder levels is mostly governed by beating of multiple excited eigenmodes, but if in a particular sample mainly localized modes are excited, the light remains confined in the vicinity of the excited guide. For stronger disorder the probability to excite such localized modes by launching light into a single site grows considerably, since the majority of modes are localized. Hence, averaged output intensity patterns are more localized at higher disorder levels. Our simulations reveal that due to repulsion from the surface for all disorder levels the number of eigenmodes localized near the boundary is smaller than those localized near the array center. The probability to encounter a localized mode for edge channel is therefore decreased compared to excitation of central guide, requiring a higher level of disorder for localization near the boundary.

To directly monitor the light evolution in particular samples, we used a fluorescence technique [18]. The diffraction patterns for the regular array are shown in Fig. 5(a,b) for central and edge excitation, respectively. The wavepacket spreading is ballistic since all eigenmodes are extended. For $S_{\rm d} = 3$ $\mu$m, the array exhibits extended as well as localized modes. Examples of delocalized light patterns are shown in Figs. 5(c,d). For these realizations the position of the input excitation was too far from the center of any localized eigenmode causing considerable spreading of light. Still, the spreading is reduced compared to the regular case. In Figs. 5(e,f), the light remains mostly localized in the vicinity of the excited site since the input was close to the center of a localized eigenmode, so that one can observe localization.

In conclusion, we observed disorder-induced wave localization at the edge of 1D photonic lattices which is weaker in the very vicinity of a surface as compared to the array center.

# References without titles

# Figure captions

Figure 1. Averaged theoretical output distributions of $\ln(I)$ for excitation of central (black curve) and edge (red curve) channels. Disorder level is $S_d = 3$ $\mu$m (a) and $S_d = 7$ $\mu$m (b).

Figure 2. Averaged experimental output intensity distributions for excitation of central (left column) and edge (right column) channels. Disorder level is $S_d = 0$ $\mu$m (a), $S_d = 2$ $\mu$m (b), and $S_d = 7$ $\mu$m (c).

Figure 3. Exponential decay rate for average output intensity distributions versus $S_d$ for excitation of edge (red symbols) and central (black symbols) waveguides. (a) Theory, (b) experiment.

Figure 4. Eigenmodes of disordered array localized close to the center (left column) and left boundary (right column) of the array. In (a), (b) $S_d = 3$ $\mu$m, while in (c), (d) $S_d = 7$ $\mu$m. Gray regions indicate positions of waveguides.

Figure 5. Fluorescence images showing propagation dynamics in regular (a),(b) and disordered arrays (c)-(f) with $S_d = 3$ $\mu$m. In (a),(c),(e) central channel is excited and in (b),(d),(f) edge channel is excited.

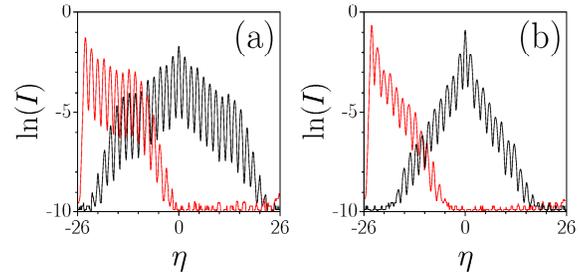

Figure 1. Averaged theoretical output distributions of $\ln(I)$ for excitation of central (black curve) and edge (red curve) channels. Disorder level is $S_{\mathrm{d}} = 3$ $\mu$m (a) and $S_{\mathrm{d}} = 7$ $\mu$m (b).

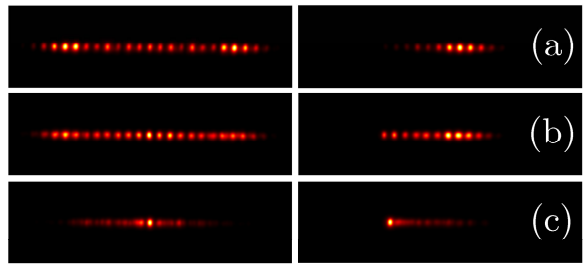

Figure 2.  Averaged experimental output intensity distributions for excitation of central (left column) and edge (right column) channels. Disorder level is $S_\mathrm{d} = 0$ $\mu$m (a), $S_\mathrm{d} = 2$ $\mu$m (b), and $S_\mathrm{d} = 7$ $\mu$m (c).

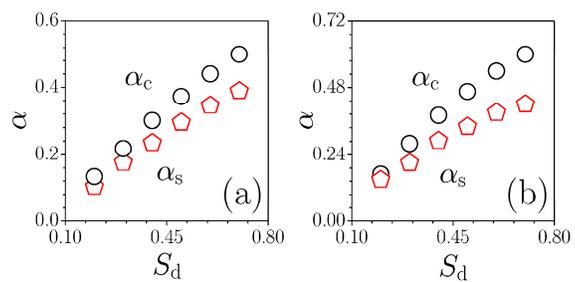

Figure 3. Exponential decay rate for average output intensity distributions versus $S_d$ for excitation of edge (red symbols) and central (black symbols) waveguides. (a) Theory, (b) experiment.

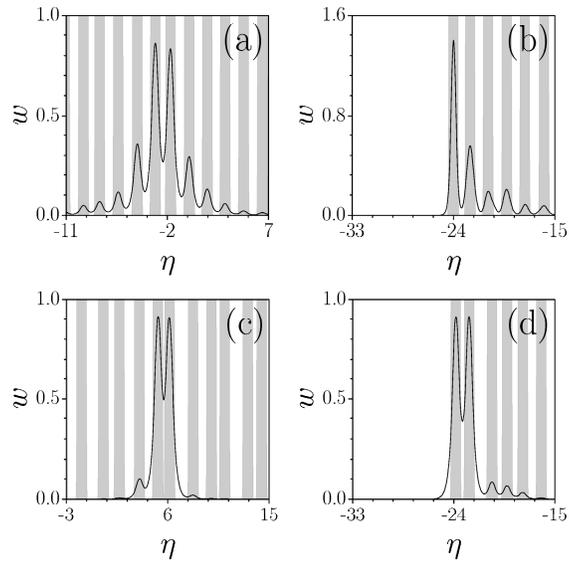

Figure 4. Eigenmodes of disordered array localized close to the center (left column) and left boundary (right column) of the array. In (a), (b) $S_\mathrm{d} = 3\ \mu\mathrm{m}$, while in (c), (d) $S_\mathrm{d} = 7\ \mu\mathrm{m}$. Gray regions indicate positions of waveguides.

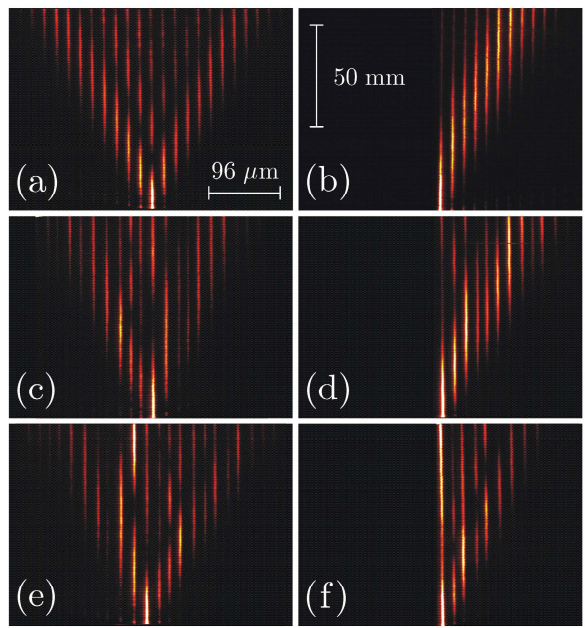

Figure 5. Fluorescence images showing propagation dynamics in regular (a),(b) and disordered arrays (c)-(f) with $S_\mathrm{d} = 3$ $\mu$m. In (a),(c),(e) central channel is excited and in (b),(d),(f) edge channel is excited.

13